\documentstyle[epsf,epsfig]{aipproc}

\newcommand{\uq}{{\sf u}}                  
\newcommand{\qq}{{\sf q}}                  
\newcommand{\qqbar}{{\sf \=q}}             
\newcommand{\gq}{{\sf g}}                  

\begin{document}


\title{$\Delta G$ from correlated high-$p_T$ hadron pairs
in polarized $l-N$ scattering}

\author{\underline{Alessandro Bravar}, Dietrich von Harrach, and Aram Kotzinian}

\address{Institut f\"ur Kernphysik, Universit\"at Mainz,
D-55099 Mainz, Germany \\}
\maketitle

\begin{abstract}
We propose to access the gluon polarization $\Delta G$
by measuring the cross section
spin-asymmetry in semi-inclusive polarized lepton -- nucleon scattering.
The photon-gluon fusion sub-process will be tagged by detecting
high-$p_T$ correlated hadron pairs in the forward hemisphere.
Selecting oppositely charged kaon pairs will allow
to suppress the background coming from gluon radiation.
\end{abstract}

\hfill

\begin{center}
Talk given at the \\
\vspace*{8pt}
Fifth International Workshop on \\
\vspace*{3pt}
{\it Deep Inelastic Scattering and QCD} \\
\vspace*{3pt}
Chicago, USA, 14--18 April, 1997 \\
\vspace*{8pt}
{\it to be published in the proceedings}
\end{center}

\newpage
\setcounter{page}{1}

\title{$\Delta G$ from correlated high-$p_T$ hadron pairs
in polarized $l-N$ scattering}

\author{\underline{Alessandro Bravar}, Dietrich von Harrach, and Aram Kotzinian}

\address{Institut f\"ur Kernphysik, Universit\"at Mainz,
D-55099 Mainz, Germany \\}
\maketitle

\begin{abstract}
We propose to access the gluon polarization $\Delta G$
by measuring the cross section
spin-asymmetry in semi-inclusive polarized lepton -- nucleon scattering.
The photon-gluon fusion sub-process will be tagged by detecting
high-$p_T$ correlated hadron pairs in the forward hemisphere.
Selecting oppositely charged kaon pairs will allow
to suppress the background coming from gluon radiation.
\end{abstract}

Polarized deep inelastic scattering (DIS) experiments have shown
that the quark spins account for only a rather small fraction of
the nucleon spin~\cite{emc}, thus implying an appreciable contribution
either from gluon spins, or possibly from orbital angular momentum.
Competing explanations exist for this result, in which the
polarized glue $\Delta G$ or negatively polarized
strange quarks $\Delta {\sf s}$ lower the quark contribution
to the nucleon spin. One way to solve this puzzle is to measure 
$\Delta G$ directly
by studying, for example, polarized semi-inclusive processes,
where the gluons enter in the initial state of the hard scattering
sub-processes at lowest order in $\alpha_s$.

\begin{figure}
\vspace{-8mm}
\begin{center}
\mbox{
\epsfxsize=14cm\epsffile{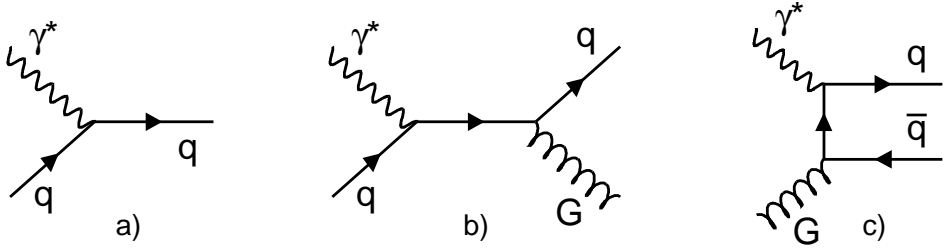}
}
\end{center}
\vspace{-5mm}
\caption{ 
Lowest order Feynman diagrams for DIS $\gamma^* N$ scattering:
a) virtual photo-absorption, b) gluon radiation (Compton diagram),
c) photon-gluon fusion (PGF).
}
\label{fig:feydia}
\end{figure}

Favorable conditions to perform such studies are given, for example,
in heavy flavor production which proceeds via photon-gluon fusion (PGF),
and in the reaction,
$\gamma^\ast \, + \, N \rightarrow 2$ high-$p_T$ jets $+ \, X$,
which goes via PGF (Fig.~\ref{fig:feydia}c)
with a non negligible background from gluon radiation
(Fig.~\ref{fig:feydia}b).
The latter would require the detection of two jets with large
transverse momenta ($p_T \, ({\rm jet}) > 5~{\rm GeV}/c$);
at moderate energies of fixed target experiments
the identification of jets is not well defined.

Alternatively to the direct jet reconstruction
we propose to look for two correlated high-$p_T$ hadrons
$h_1$ and $h_2$ in the forward hemisphere ($x_F > 0$)
with $p_T (h_1) > p_{T,min}$ and $h_2$
opposite in azimuth to $h_1$ with $p_T (h_2) > p_{T,min}$.
The measured cross section spin-asymmetry for $h_1 + h_2$ production
can be related to the gluon polarization $\Delta G$.
The idea is based on the observation that hadron
transverse momentum spectra
get significant contributions from the first order QCD diagrams
already at relatively low values of $p_T$
($p_T > 1~{\rm GeV}/c$)~\cite{e665}.
In events with an underlying $2 \rightarrow 2$ hard scattering QCD sub-process
(Fig.~\ref{fig:feydia}b \qq\gq-event and Fig.~\ref{fig:feydia}c \qq\qqbar-event)
the two outgoing partons will fragment to two high-$p_T$ hadrons 
with $x_F > 0$ and,
on average, larger transverse momenta than hadrons produced in events
with a leading order QED sub-process (Fig.~\ref{fig:feydia}a \qq-event).
From the numerical analysis based on Monte Carlo studies a
$p_{T,min} > 1.0 - 1.5 ~{\rm GeV}/c$ is found to be sufficiently large
to suppress almost completely the QED sub-process and to tag
effectively the PGF process.
A more detailed discussion will be presented elsewhere~\cite{BHK97}.

The parton distribution functions (PDF) are probed in 
\qq\gq- and \qq\qqbar-events at a momentum fraction
$\eta = (\hat{s} + Q^2)/2 M E_{\gamma^\ast} = x_{Bj} \, (\hat{s}/Q^2 +1)$,
where ${\hat s}$ is the c.m. energy of the hard sub-process and
$E_{\gamma^\ast}$ the energy of the virtual photon.
In fixed target experiments the lepton energies are between
$ 100 - 500~{\rm GeV}$,
which corresponds a c.m. energy of $\sqrt{s} = 15 - 30$~GeV for the
$\gamma^\ast - N$ system.
To have reasonably large values of $\hat{s}$,
one can probe the PDF only above $\eta \sim 0.05$
($\langle \eta \rangle \sim 0.1$).
As far as the spin of the nucleon is concerned, this appears to be the most
interesting region, since the largest contribution to $\Delta G$
is expected to come from that region~\cite{gs96}.
The proposed method will probe the PDF at a scale
around $10~{\rm GeV}^2$.

The following selection criteria are found to enhance the contributions
of the \qq\qqbar- and \qq\gq-events to the correlated hadron pair
production cross section
and to suppress almost completely the contribution of the \qq-events.

\noindent {\bf A} -- $p_T$ {\it cut:}
There should be two hadrons in the event
with $p_T > 1.0 - 1.5~{\rm GeV}/c$.
The $p_T$ cut will depend also on the final event yields and
a higher $p_T$ cut is preferable.

\noindent {\bf B} -- $x_F$ {\it cut:}
To avoid fragmentation effects from the target remnant
only hadrons produced in the forward hemisphere
($x_F>0$) should be considered.
We also require $z > 0.1$ for each hadron.

\noindent {\bf C} -- $\Delta \phi$ {\it cut:}
The two hadrons should be found opposite in azimuth,
such that $150^\circ < \Delta \phi < 210^\circ$.

\noindent {\bf D} -- {\it mass cut:}
To avoid the divergences in the QCD matrix elements in addition to the
$p_T$ cut, we require the invariant mass of the hadron pair $m(h_1,h_2)$
to exceed 2 or 3~GeV/$c^2$ (${\hat s} > m^2(h_1,h_2)$).
A higher mass cut is preferable.

\begin{figure}
\vspace{-5mm}
\begin{center}
\mbox{
\epsfxsize=14cm\epsffile{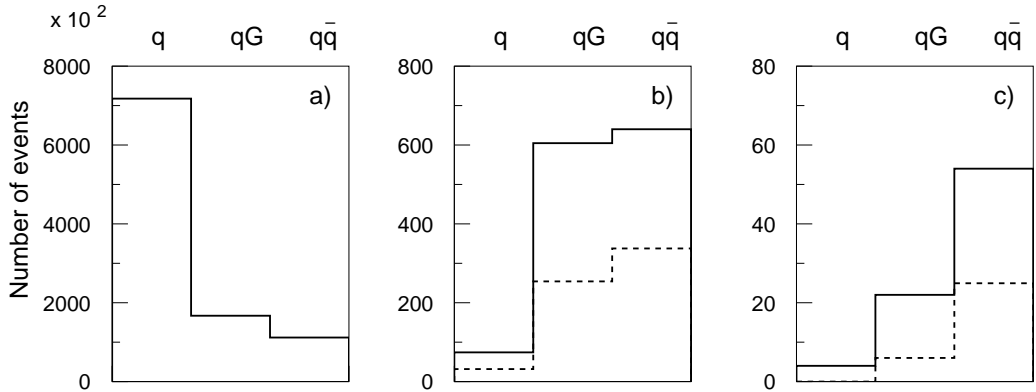}
}
\end{center}
\caption{ 
Contributions of different sub-processes to the cross section:
(a) no cuts, (b) opposite charged hadron pairs,
(c) opposite charged kaon pairs
($p_T > 1~{\rm GeV}/c$ and $m(h_1,h_2) > 3~{\rm GeV}/c^2$).
The dashed line shows the effect of the asymmetric $x_F$ cut.
The event yields are normalized to $10^6$ generated events.
}
\label{fig:yields}
\end{figure}

Two competing processes of the same order in $\alpha_s$ contribute to the
cross section; however only the PGF is of interest for the extraction
of $\Delta G$, while the Compton diagram acts as a background.
Around $\eta \sim 0.1$ most of the Compton background
comes from scattering off \uq~quarks.
The relative contribution of the PGF over the Compton diagram
can be further enhanced from the following considerations.

\noindent {\bf E} -- {\it oppositely charged hadrons:}
Fragmenting partons in \qq\qqbar-events have opposite charges
and favored fragmentation will lead to opposite charged hadrons in contrast
to \qq\gq-events, where the gluon fragments with equal probability
to positive and negative hadrons.

\noindent {\bf F} -- {\it $K^+K^-$ pairs:}
The production of strange hadrons in fragmentation
is quite suppressed compared to non-strange hadrons, unless there is
already an {\sf s}~quark in the initial state.
This effect is parametrized with a strangeness suppression factor $\gamma_s$,
which presently ranges between 0.2 and 0.3.
The production of a high $p_T$ correlated kaon pair will be 
strongly suppressed,
unless there is already a fragmenting ${\sf s}{\sf \bar{s}}$ pair
which can be produced via the PGF only.

\noindent {\bf G} -- {\it asymmetric $x_F$ cut:}
The Compton process in the partonic c.m. is peaked in the backward direction,
while the PGF has a symmetric distribution.
In the hadronic c.m. the first process will generate faster
positive hadrons, because of the favored fragmentation of \uq~quarks
to positive hadrons.
Thus by requiring additionally $x_F^+ < x_F^-$ the relative contribution
of the PGF can be further enhanced.

We have performed a numerical analysis in the kinematical conditions of the
COMPASS experiment~\cite{comp} with a 200~GeV/$c$ $\mu^+$
beam incident on a deuteron (iso-scalar) target
and the following kinematical requirements: 
$0.5 < y_{Bj} < 0.9$, $W^2 > 200~{\rm GeV}^2/c^2$, and $Q^2 > 1~{\rm GeV}^2$.
A slightly modified version of the LEPTO~\cite{lepto} event generator
has been used with the GRV94LO unpolarized parton densities.

Figures~\ref{fig:yields}b and~\ref{fig:yields}c show the relative suppression
of the \qq-events for oppositely charged hadron and kaon pairs,
respectively,
with $p_T > 1.0~{\rm GeV}/c$ and $m(h_1,h_2) > 3~{\rm GeV}/c^2$
compared to Fig.~\ref{fig:yields}a
where no selections on the final hadronic state have been applied.
The fraction of \qq-events in the selected samples is about 5~\%.
Let's define $R = \sigma^{PGF} / \sigma^{Compt}$;
$R \stackrel {>}{\sim} 1$ for the selected hadron pairs
and $R \stackrel {>}\sim 2$ for the kaon pairs.
The reduction factor for the hadron pairs is $\sim 1.5 \times 10^{-3}$
and for the kaon pairs is $\sim 0.8 \times 10^{-4}$.

Including the photo-production $Q^2 \rightarrow 0$ limit
will significantly increase the event yields.
The relevant scales are set by $\hat{s}$ and/or $p_T^\ast$
and are around 10~GeV$^2$, and the possible
contribution from non-pointlike photons is expected to be very small.
We estimate the total non-diffractive cross section for producing such
correlated hadron pairs to be around $150 \pm 50~{\rm nb}$,
by extrapolating from the DIS cross section to $Q^2 \rightarrow 0$
using a dipole form factor with a mass parameter of 10~GeV$^2$
set by the hard process scale.

\begin{figure}
\vspace{-5mm}
\begin{center}
\mbox{
\epsfxsize=14cm\epsffile{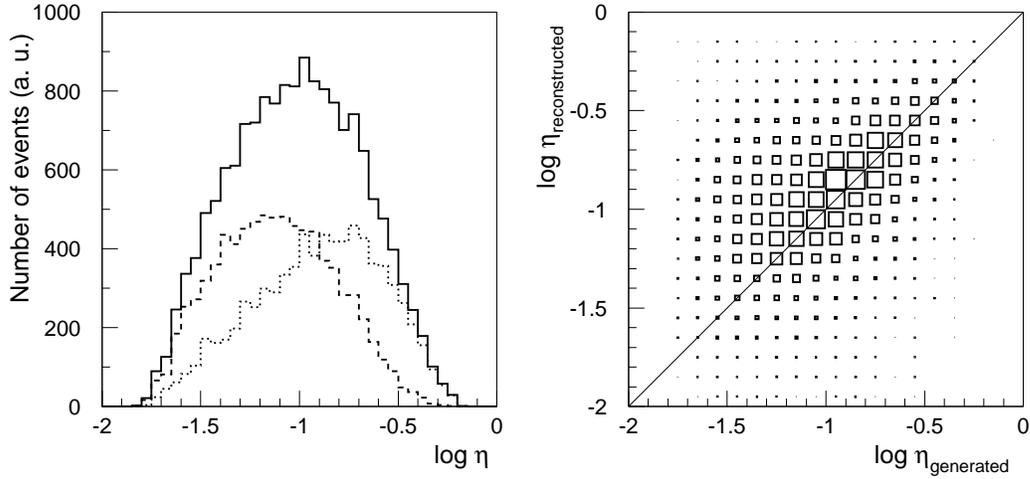}
}
\end{center}
\vspace{-5mm}
\caption{ 
{\it left:} Gluon $\eta$ spectra for \qq\qqbar-events (dashed line)
and quark $\eta$ spectra for \qq\gq-events (dotted line) for accepted
high-$p_T$ hadron pairs. The full line is the sum of the two. 
{\it right:} Correlation between the generated and the reconstructed 
parton momentum fraction $\eta$.
}
\label{fig:gluon}
\end{figure}

Figure~\ref{fig:gluon} shows the $\eta$ distribution for
the accepted hadron pairs.
Also shown is the correlation between the generated and the reconstructed
$\eta$ using the kinematics of the two selected hadrons only.
The $\eta$ distribution is peaked around $0.1$,
and above $\eta \sim 0.15 - 0.20$ the
\qq\gq-events dominate over the \qq\qqbar-events.
The gluon $\eta$ region accessible via the hadron pairs with a 200~GeV beam 
ranges from $\eta \sim 0.04$ to $\sim 0.2$.
The correlation in $\eta$ allows to study $\frac{\Delta G}{G} (\eta)$
and further suppress the Compton background. 

The cross section spin-asymmetry $A_{LL}^{lN \rightarrow h_1 h_2}$
can be approximated over a small phase space as
\begin{equation}
A_{LL}^{lN \rightarrow h_1 h_2} 
\sim 
\langle \hat{a}_{LL}^{\gamma^\ast G \rightarrow q\bar{q}}\rangle
\frac{\Delta G}{G} \, \frac{R}{1+R} + 
\langle\hat{a}_{LL}^{\gamma^\ast q \rightarrow qG}\rangle
\frac{\Delta u}{u} \, \frac{1}{1+R} \; .
\label{eq:all}
\end{equation}
The $\langle\hat{a}_{LL}\rangle$'s are the scattering asymmetries
at partonic level.
The small contribution to $A_{LL}^{lN \rightarrow h_1 h_2}$
from \qq-events has been neglected.
In order to extract $\Delta G / G$ from (1)
the quark polarizations $\Delta {\sf q} / {\sf q}$ and the ratio $R$ must be
known.
The first can be taken from other measurements, while the latter can
be estimated with an event generator.
Since $\langle\hat{a}_{LL}^{\gamma^\ast+G \rightarrow q+\bar{q}}\rangle \sim -1$
and $\langle\hat{a}_{LL}^{\gamma^\ast+q \rightarrow q+G}\rangle \sim +0.5$
the contribution of the Compton sub-process to
the asymmetry is further suppressed.

Using the polarized parton densities of~\cite{gs96},
we have estimated a value for the photon nucleon asymmetry
$A_{LL}^{\gamma N \rightarrow h^+h^-}$ of about $-12~\%$, 
and a value for $A_{LL}^{\gamma N \rightarrow K^+K^-}$ of about $-20~\%$
around $\eta \sim 0.1$.
Note that these asymmetries have opposite sign compared to the
polarized open charm muoproduction asymmetry~\cite{comp}.

In a high luminosity fixed target experiment like COMPASS~\cite{comp},
integrated luminosities $L$ of 1 -- 2~fb$^{-1}$
can be easily achieved in one year.
With $L = 2~{\rm fb}^{-1}$ and a cross section of 150~nb,
about 60~k $K^+K^-$ pairs with
$p_T > 1~{\rm GeV}/c$ and $m(K^+,K^-) > 2~{\rm GeV}/c^2$, and
about 80~k $h^+h^-$ pairs with
$p_T > 1.5~{\rm GeV}/c$ and $m(h^+,h^-) > 3~{\rm GeV}/c^2$
can be collected in one year of data taking.
With a beam polarization of $\sim 80~\%$ and an effective target polarization
of $\sim 25~\%$ per nucleon,
these event samples will give a statistical precision of about $2 - 3~\%$ 
for both $A_{LL}^{\gamma N \rightarrow h^+h^-}$ and
$A_{LL}^{\gamma N \rightarrow K^+K^-}$.

We have estimated a statistical precision on $\Delta G / G$ of about 5~\%
for kaon pairs and of about 8~\% for hadron pairs.
The precision on the extraction of $\Delta G / G$ is mostly
affected by the Compton {\it background} subtraction,
which is larger in the $h^+h^-$ channel compared to the $K^+K^-$ one.
Combining both channels might yield a better determination of 
$\Delta G / G$ because of the different backgrounds involved,
while the gluon densities remain the same.
The asymmetric $x_F$ cut has not been used for these estimates.
Adding this cut will reduce the systematic uncertainty in the
background subtraction.



\begin{thebibliography}{99}

\bibitem{emc}
EM~Collaboration, J.~Ashman {\it et al.}, Nucl. Phys. B {\bf 328}, 1 (1989). 

\bibitem{e665}
E665~Collaboration, M.R.~Adams {\it et al.},
Phys. Rev. D {\bf 48}, 5057 (1993); \\
EM~Collaboration, M.~Arneodo {\it et al.}, Z. Phys. C {\bf 36}, 527 (1987).

\bibitem{BHK97}
A.~Bravar {\it et al.}, work in preparation.

\bibitem{gs96}
T.~Gehrmann and W.J.~Stirling, Phys. Rev. D {\bf 53}, 6100 (1996). 

\bibitem{comp}
The COMPASS Collaboration, COMPASS Proposal, CERN/SPSLC 96-14, SPSC/P297,
March 1996; \\
A.~Bravar, these proceedings.

\bibitem{lepto}
G.~Ingelman, A.~Edin, and J.~Rathsman,
Comp. Phys. Comm. {\bf 101}, 108 (1997).

\end{thebibliography}
\end{document}